\documentclass[aps,pra,twocolumn,showpacs,superscriptaddress]{revtex4-2}
\usepackage{amsfonts,mathrsfs,amsmath,amsthm,amssymb,bbold}
\usepackage{hyperref}
\usepackage{epsfig}
\usepackage{bm,mathrsfs}
\usepackage{lineno,hyperref}
\usepackage[margin=25mm]{geometry} 
\usepackage{array} 
\usepackage{color}
\usepackage{xcolor,colortbl} 
\usepackage{soul}
\usepackage{afterpage}
\usepackage{capt-of}
\usepackage{makecell}
\usepackage{graphicx}
\usepackage{ulem}
\usepackage{cancel}
\usepackage{tabularx}
\usepackage{siunitx}
\usepackage[version=4]{mhchem}
\DeclareSIUnit\gauss{Gauss}
\DeclareSIUnit\muB{\ensuremath{\mu_{\mathrm{B}}}}
\DeclareSIUnit\oersted{Oe} 
 
\setcellgapes{4pt}
\newcolumntype{P}[1]{>{\arraybackslash}p{#1}}

\newcommand*{\sm}{\textrm{Supplementary Material}}

\newcommand{\etal}{\textit{et al.}}
\newcommand{\ie}{\textit{i.e.}}

\begin{document}  
\title{Native defects and $p$-type dopability in transparent $\beta$-TeO$_2$: A first-principles study}
\author{Vu Thi Ngoc Huyen}
\author{Soungmin Bae}
\author{Rafael Costa-Amaral}
\author{Yu Kumagai}
\thanks{yukumagai@tohoku.ac.jp}
\affiliation{Institute for Materials Research, Tohoku University, Sendai, Miyagi 980--8577, Japan}

\date{\today}

\begin{abstract}
Although $\beta$-\ce{TeO2} is a promising $p$-type transparent conducting oxide (TCO) due to the large optical gap
($\sim$ 3.7 eV) and a light effective hole mass, its hole dopability still remains unexplored.
In this work, electronic structure of $\beta$-\ce{TeO2} and its point defects are investigated using the HSEsol functional with the band-gap-tuned mixing parameter.
Our calculations reveal that $\beta$-\ce{TeO2} exhibits a significant difference between
the fundamental and optical band gaps because lower energy optical transitions are dipole forbidden.
Additionally, it has a low hole effective mass, especially in-plane.
The point defect calculations show that $\beta$-\ce{TeO2} is intrinsically an insulator.
From systematic calculations of the trivalent dopants as well as hydrogen, Bi doping is suggested as the best candidate as an acceptor dopant.
This work paves the way for the material design of the $p$-type $\beta$-\ce{TeO2}.
\end{abstract}
\maketitle
\section{Introduction}\label{sec:intro}
%
\begin{figure} 
\centering 
\includegraphics[width=8cm]{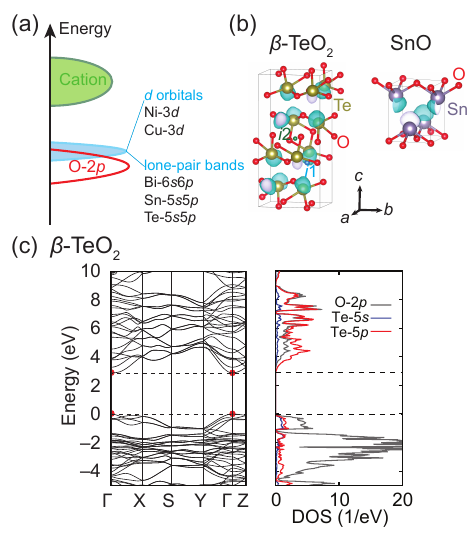}\\
\captionof{figure}{
(a) A schematic energy diagram illustrating strategies to raise the
    valence band maximum in oxides for introducing hole carriers.
(b) Crystal structures of $\beta$-TeO$_2$ and SnO with electron
    localization functions (cyan) that indicate lone-pair electrons.
    The isosurfaces are set to 90\% of the maxima in each case.
(c) Electronic band structure and projected density of states in $\beta$-TeO$_2$.
    The red circles in the band structure indicate the valence band maximum
    and the conduction band minimum.
    See the text for computational details.
}
\label{fig:concept}
\end{figure}
The $p$--$n$ junction of transparent conducting oxides (TCOs) would be
crucial for wide range of applications such as displays~\cite{afre2018transparent},
bipolar transistors~\cite{takeshi2017chemical},
photovoltaics~\cite{al2022photovoltaic},
and inverter circuits~\cite{shijeesh2020complementary}.
Despite the availability of $n$-type TCOs,
such as Sn-doped In$_2$O$_3$~\cite{kim2020persistent} and
Al-doped ZnO~\cite{majumder2003investigations},
$p$-type counterparts remain scarce due to the difficulty of hole doping
and/or low hole mobility~\cite{PhysRevApplied.19.034063,PhysRevMaterials.3.044603}.

The main obstacle to $p$-type doping in oxides originates from
their electronic structures,
where the valence band maxima (VBMs) consisting of O-$2p$ states are
too deep to introduce hole carriers.
Therefore, the strategy for designing $p$-type TCOs is
to place the orbitals with higher energy levels above the O-$2p$ bands
\cite{kawazoe1997p}, as illustrated in Fig.~\ref{fig:concept}(a).
Such strategy has been demonstrated with Ni-$3d$ orbitals
in NiO~\cite{molaei2013crystallographic},
Cu-$3d$ orbitals in CuAlO$_2$~\cite{kawazoe1997p,gake2021point}
and CuGaO$_2$~\cite{gake2021point,ueda2001epitaxial},
Sn-$5s$ and $5p$ orbitals in SnO~\cite{varley2013ambipolar}
and K$_2$SnO$_2$~\cite{hautier2013identification},
and Bi-$6s$ and $6p$ orbitals in BaBiO$_3$~\cite{schoop2013lone}.
When additional orbitals are close to the O-$2p$ orbitals in energy,
they typically form covalent bonds with the O-$2p$ bands.
This results in another benefit: more dispersive valence bands with reduced hole effective masses.

Regarding SnO, although it shows the $p$-type conductivity with a small hole effective mass~\cite{allen2013understanding},
the optical band gap is 2.70~eV,
which is slightly below the threshold to achieve transparency in the visible range.
This motivates us to explore $p$-type TCO candidates
with similar lone pair characteristics as Sn$^{2+}$ but with a higher optical gap.

\ce{TeO2} has been suggested as a possible candidate in the same manner as SnO
because Te$^{4+}$ has the same electronic configuration as Sn$^{2+}$.
We display electron localization functions~\cite{silvi1994classification} of $\beta$-\ce{TeO2} and \ce{SnO}
in Fig.~\ref{fig:concept}(b), both of which exhibit
lone-pair orbitals composed of the mixture of $5s$ and $5p$ orbitals.
The density of states (DOS) in $\beta$-\ce{TeO2} [Fig.~\ref{fig:concept}(c)] shows
that the VBM is lifted up by the presence of Te 5$s$ and 5$p$ lone pairs
that hybridize with the O-$2p$ bands.
The $p$-type behavior has been reported
for a two-dimensional (2D) bilayer $\beta$-\ce{TeO2}
with high hole mobility and high band gap around 3.7 eV~\cite{zavabeti2021high}.
\textit{Ab initio} calculations and subsequent experiments
have also confirmed its high hole mobility
and small hole effective masses~\cite{shi2023electronic,PhysRevApplied.17.064010,nguyen2024stereochemically,roginskii2017comparative,suehara2004non,C8NR01028E}.

According to the material properties,
the three-dimensional (3D) $\beta$-\ce{TeO2} is expected to be a good candidate for a $p$-type TCO
due to its large optical band gap and small hole effective mass~\cite{shi2023electronic}.
Moreover, in view of the defect perspectives,
since Te and O have very different ionic radii (0.66 and 1.35 $\mathrm{\AA}$ in the 4-coordination~\cite{shannon1976revised}),
the formation of antisite-type defects that often
cause the Fermi-level pinning within the band gap are expected to be prohibitive.
However, the experimental report of the $p$-type conduction is still limited to a
2D field-effect transistor (FET) $\beta$-\ce{TeO2}~\cite{zavabeti2021high}.

To achieve fundamental insight of $p$-type behavior of 3D $\beta$-\ce{TeO2},
we investigate the intrinsic defects and $p$-type dopants
using the HSEsol hybrid functional~\cite{Schimka2011,heyd2003hybrid,ge2006erratum,krukau2006influence}.
We first discuss the electronic structures of the TeO$_2$ polymorphs.
We have found that only $\beta$-\ce{TeO2} exhibits a significant difference between
the fundamental gap and the optical band gap
because the lower energy optical transitions are dipole forbidden.
For the calculations of the intrinsic defects in $\beta$-\ce{TeO2},
vacancies, antisites, and interstitials were considered and equilibrium Fermi levels and carrier concentrations were estimated.
Trivalent acceptor dopants as well as hydrogen impurities were also calculated.
Consequently, Bi was suggested as the best $p$-type dopant exhibiting the $(+/-)$ acceptor transition around 0.47~eV.
Furthermore, we propose to dope acceptors in high concentrations to reduce the acceptor levels.

Initially, the computational details are explained in Sec.~\ref{sec:com}.
The electronic structures and optical properties of the \ce{TeO2} polymorphs
are then discussed in Sec.~\ref{sec:polymorphs}.
In Secs.~\ref{subsec:native-point-defects} and~\ref{sec:dopants},
the formation energies of intrinsic defects and impurities in $\beta$-\ce{TeO2} are discussed, respectively.
Finally, the conclusion is summarized in Sec.~\ref{sec:conclu}.

\section{Computational details}\label{sec:com}
First-principles calculations were performed
using the projector augmented-wave (PAW) method~\cite{blochl1994projector,kresse1999ultrasoft},
as implemented in VASP~\cite{PhysRevB.54.11169}.
Te-$5s$ and $5p$, and O-$2p$ were then described as valence electrons.
Information on the PAW datasets
and the valence electrons of the other elements are provided in the \sm.
The HSEsol functional, \ie, the combination of
the Heyd-Scuseria-Ernzerhof (HSE) hybrid functional
with the PBEsol functional was employed~\cite{Schimka2011}.
The screening distance is then fixed at 0.2 $\text{\AA}^{-1}$,
and the Fock exchange mixing parameter is tuned to $\alpha^\text{HSE}=0.175$ to reproduce
the experimental optical band gap of $\beta$-\ce{TeO2} [see Sec.~\ref{sec:results}(a)].
The plane-wave cutoff energy was set to 520 eV for structure optimization
to reduce the Pulay stress and to 400 eV for the other calculations.

For the structure optimization and band structure calculations,
4$\times$4$\times$3, 3$\times$3$\times$2 and 4$\times$4$\times$2 $k$-point grids
were used for the $\alpha$, $\beta$, and $\gamma$ phases, respectively.
For metals and insulators that were calculated for constructing the chemical potential diagrams (CPDs),
the $k$-point densities were set to 2.5 and 5.0$\textup{~\AA}^{-1}$, respectively.
The force and stress convergence criteria were
set to 5 $\textup{meV/\AA}$ and 0.08 $\textup{GPa}$, respectively.
The effective mass tensors, DOS, and dielectric functions
were calculated with tripled $k$-points along all three directions
to enhance the accuracy.
The static dielectric tensors ($\epsilon_{0}$) were calculated
from the sum of the ion-clamped ($\epsilon_{\infty}$) and
ionic dielectric contributions ($\epsilon_{\rm ion}$)
based on the density functional perturbation theory~\cite{PhysRevB.33.7017,PhysRevB.73.045112,PhysRevB.63.155107}.
The $\epsilon_{\infty}$ were obtained from the HSEsol calculations,
whereas the $\epsilon_{\rm ion}$ were acquired from the PBEsol calculations
to reduce the computational costs.
For the $\epsilon_{\infty}$, the local field effects are included.

The band-averaged effective mass tensors were calculated using BoltzTraP2~\cite{madsen2018boltztrap2},
with the carrier concentration and temperature set
to $10^{16} \text{cm}^{-3}$ and 300$~\mathrm{K}$, respectively.
For the calculations of the optical absorption spectra,
the real parts of the dielectric functions were derived from the imaginary parts
via the Kramers-Kronig transformation with a complex shift of 0.01~eV\@~\cite{madsen2018boltztrap2}.
The optical band gaps were determined from the Tauc plots
to maintain consistency with experimental optical measurements~\cite{shi2023electronic}.

The band paths were automatically set using SeeK-path~\cite{hinuma2017band}.
The irreducible representations of the electronic bands at the $\Gamma$ point
were determined using the IrRep code~\cite{IRAOLA2022108226,elcoro2017double}.

For the defect calculations, we adopted a 2$\times$2$\times$1 supercell of the conventional unit cell of $\beta$-\ce{TeO2} with 96 atoms and $\alpha$-\ce{TeO2} with 48 atoms. 
A 2$\times$2$\times$1 Monkhorst-Pack $k$-point mesh~\cite{PhysRevB.13.5188} was used for $\beta$--\ce{TeO2},
and a 2$\times$2$\times$2 mesh was used for $\alpha$-\ce{TeO2}.
The internal atomic positions were relaxed until the residual forces
were reduced to less than 30 $\mathrm{meV/\AA}$ with fixed theoretical lattice constants.
In all the defect calculations, spin polarization was allowed.

We constructed the CPDs using the total energies, ignoring vibrational and entropy contributions.
In this study, we retrieved the candidate competing phases
that are either stable or slightly unstable by 0.01 meV/atom  from the Materials Project~\cite{jain2013commentary}.
For the standard state of O, we calculated an \ce{O2} molecule in the spin triplet configuration.
See also the \sm\ for the list of the competing phases considered in this study.

We used the VISE code (version 0.6.6) to generate all the VASP input settings,
and PYDEFECT~\cite{PhysRevMaterials.5.123803} to generate the defect models and
to analyze the first-principles calculation results.
The formula to calculate the formation energy of defect $D^q$
in charge $q$ ($E_f[D^{q}]$) is described elsewhere~\cite{Oba_2018}.
We corrected the defect formation energies of charged defects
using the extended FNV method as described in~\cite{PhysRevLett.102.016402,PhysRevB.89.195205,PhysRevB.90.125202}.

The effective $U$ value between $q-1$, $q$, $q+1$ is evaluated as~\cite{PRXEnergy.2.043002}
\begin{equation}
    U^\text{eff}(q-1/q/q+1) = E_f[D^{q-1}] + E_f[D^{q+1}] - 2E_f[D^{q}].
    \label{eq:equation}
\end{equation}

The carrier concentrations as a function of the Fermi level were calculated
from the DOS via the Fermi-Dirac distribution~\cite{PhysRevB.90.125202}.
The defect concentrations were estimated from the Boltzmann distribution,
considering the site degeneracies that depend on the defect site symmetries in the relaxed structures
and the spin degeneracies~\cite{PhysRevB.90.125202}.

\section{Results and discussion}\label{sec:results}
\subsection{Fundamental properties of TeO$_2$ polymorphs}\label{sec:polymorphs}
\begin{figure}
\centering 
\includegraphics[width=8cm]{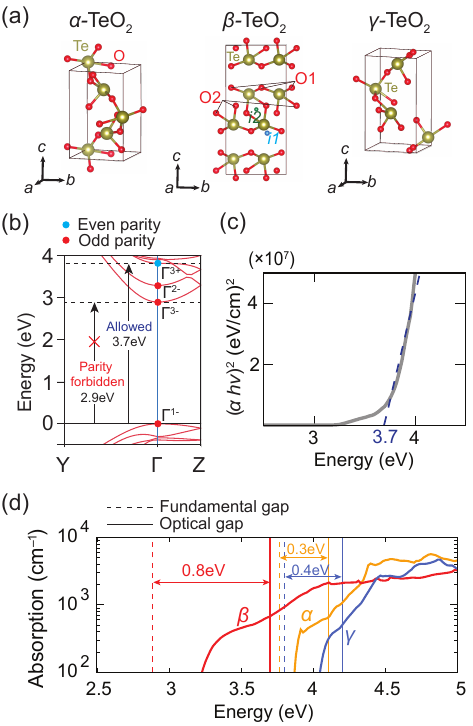}\\
\captionof{figure}{
(a) Conventional unit cells of three TeO$_2$ polymorphs;
    the interstitial sites $i_1$ and $i_2$ in $\beta$-TeO$_2$ are shown
    as small cyan and green spheres, respectively.
    Note that Fig.~\ref{fig:concept}(b) provides another angle view for $\beta$-TeO$_2$.
(b) Electronic band structure near the band edges in $\beta$-TeO$_2$.
    From the irreducible representations,
    the dipole forbidden and dipole allowed transitions at the $\Gamma$ point are described.
(c) Tauc plot for the $\beta$-TeO$_2$ phase.
    The fitting range is set to 3.8--4.0 eV, as in the experiment~\cite{shi2023electronic}.
    See text for details.
(d) Optical absorption spectra of $\alpha$-, $\beta$-, and $\gamma$-TeO$_2$ phases.
    The fundamental and optical band gaps are also depicted with dashed and solid vertical lines, respectively.
}
\label{fig:3phase}
\end{figure}
\begin{table*}
\captionof{table}{
    The experimental (Exp.)~\cite{P_A_Thomas_1988,BEYER+1967+228+237,CHAMPARNAUDMESJARD20001499} and calculated (Calc.) lattice constants,
    calculated hole effective masses ($m_p^*$), and calculated ion-clamped and ionic
    dielectric constants for three TeO$_2$ polymorphs.}
\label{tab:lat}
\begin{center}
\begin{tabular}{P{1.5cm}P{1.8cm}P{2.0cm}P{1.5cm}P{1.5cm}P{2.0cm}P{2.2cm}P{1.5cm}}
\hline
\hline
& &\multicolumn{3}{c} {Lattice constants} & &\multicolumn{2}{c} {Dielectric constants}\\
\cline{3-5}  \cline{7-8}
Phase & Direction & Exp. (\AA) & Calc. (\AA) & Diff($\%$) & $m_p^*(m_0)$  & Ion-clamped & Ionic \\
\hline
$\alpha$&$a$ & 4.808 &4.78 & $-0.52$& 1.85& 4.9 & 18.8 \\
        &$c$ & 7.612 &7.49 & $-1.62$&3.69 & 5.6 & 20.5 \\
\hline
$\beta$ &$a$ & 5.464 &5.30 & $-2.98$& 0.75& 4.7 & 22.1 \\
        &$b$ & 5.607 &5.62 & $+0.31$& 0.74& 5.5 & 18.4 \\
        &$c$ &12.035 &12.07& $+0.28$& 2.23& 4.2 &  4.7\\
\hline
$\gamma$&$a$ & 4.898 & 4.84& $-1.20$&6.88 & 4.6 & 9.2 \\
        &$b$ & 4.351 & 4.32& $-0.71$&3.35 & 4.4 & 9.0 \\
        &$c$ & 8.576 & 8.64& $+0.76$&2.31 & 4.9 & 15.4 \\
\hline
\hline
\end{tabular}
\end{center}
\end{table*}
\ce{TeO2} has three polymorphs:
$\alpha$-\ce{TeO2} $(P4_1 2_1 2)$~\cite{P_A_Thomas_1988}, $\beta$-\ce{TeO2} $(Pbca)$~\cite{BEYER+1967+228+237},
and $\gamma$-\ce{TeO2} $(2_1 2_1 2_1)$~\cite{CHAMPARNAUDMESJARD20001499}. 
As illustrated in Figs.~\ref{fig:3phase}(a), each Te atom is surrounded
by four O atoms, forming a \ce{TeO4} disphenoid network.
Different from $\alpha$- and $\gamma$-\ce{TeO2} phases,
$\beta$-\ce{TeO2} has a 2D layered structure in the $a$--$b$ plane with the \ce{TeO4} network
being disconnected along the $c$-direction.

As demonstrated for $\beta$-\ce{TeO2} in Fig.~\ref{fig:concept}(c),
the valence bands in three \ce{TeO2} phases are mainly formed from O-$2p$ states
hybridized with Te-5$s$ and Te-5$p$ states.
The electronic structures of the $\alpha$- and $\gamma$-\ce{TeO2} phases are also provided in the \sm.
Consequently, the VBM of $\beta$-\ce{TeO2} is raised
in energy from the top of the O-$2p$ bands by about 1.5~eV,
which can make $p$-type doping easier, as discussed in Sec.~\ref{sec:intro}.
On the other hand, the conduction band is primarily composed of the Te-$5p$ orbitals.

Here, we determine the optical band gaps from the Tauc plots,
to keep the consistency with experiment~\cite{shi2023electronic}.
Figure~\ref{fig:3phase}(c) shows $(\alpha h \nu)^2$ against $h \nu$;
its linear extrapolation is used to determine the direct-allowed dipole transition gap.
As shown, the $x$-intercept exactly matches the experimental value of 3.7 eV,
indicating the adequacy of $\alpha^\text{HSE}=0.175$.

The calculated fundamental \textit{direct} gap with $\alpha^\text{HSE}=0.175$
is 2.9 eV, much smaller than the optical gap.
This is primarily because the transitions from the VBM located at the $\Gamma$ point
to the lowest and second-lowest unoccupied states at the $\Gamma$ point
are dipole forbidden as shown in Fig.~\ref{fig:3phase}(b).

Such a smaller fundamental gap is advantageous for doping in general,
as exemplified by In$_2$O$_3$~\cite{PhysRevLett.100.167402}.
Using the same mixing parameter,
we obtain fundamental gaps of 3.8 eV for both $\alpha$- and $\gamma$-TeO$_2$,
with optical gaps of 4.1 and 4.2 eV, respectively.
Therefore,  the smaller fundamental gap is unique to the $\beta$ phase
[see also their optical absorption spectra in Fig.~\ref{fig:3phase}(d)]
\footnote{Ref.~\cite{nguyen2024stereochemically} has reported the optical absorption spectrum of $\beta$-\ce{TeO2}, but their calculated absorbance
differs from ours by about a factor of 100.}.

Table~\ref{tab:lat} summarizes the calculated lattice constants
alongside the available experimental values~\cite{P_A_Thomas_1988,BEYER+1967+228+237,CHAMPARNAUDMESJARD20001499}.
The differences are less than $1.6\%$ in most cases:
an exception is the $a$ lattice constant of $\beta$-TeO$_2$, which is underestimated by about 3\%.
Note that, however, we have confirmed that the calculated band gap at the experimental lattice constants is the same as that at the theoretical lattice constants.

Regarding the hole effective masses ($m_p^*$),
$\beta$-TeO$_2$ generally exhibits smaller $m_p^*$ compared to the other phases.
Notably, $m_p^*$ in the $a$--$b$ plane is only 0.74--0.75 $m_0$,
where $m_0$ is the free-electron rest mass.
This is comparable to the calculated value of
0.84 and 1.06 $m_0$ along the X--$\Gamma$ and Y--$\Gamma$ directions by Shi \etal~\cite{shi2023electronic},
and that for the bilayer of $\beta$-\ce{TeO2} (0.51 $m_0$) by Zavabeti \etal~\cite{zavabeti2021high}.
%
\subsection{Native point defects}\label{subsec:native-point-defects}
\begin{figure}
\centering
\includegraphics[width=8cm]{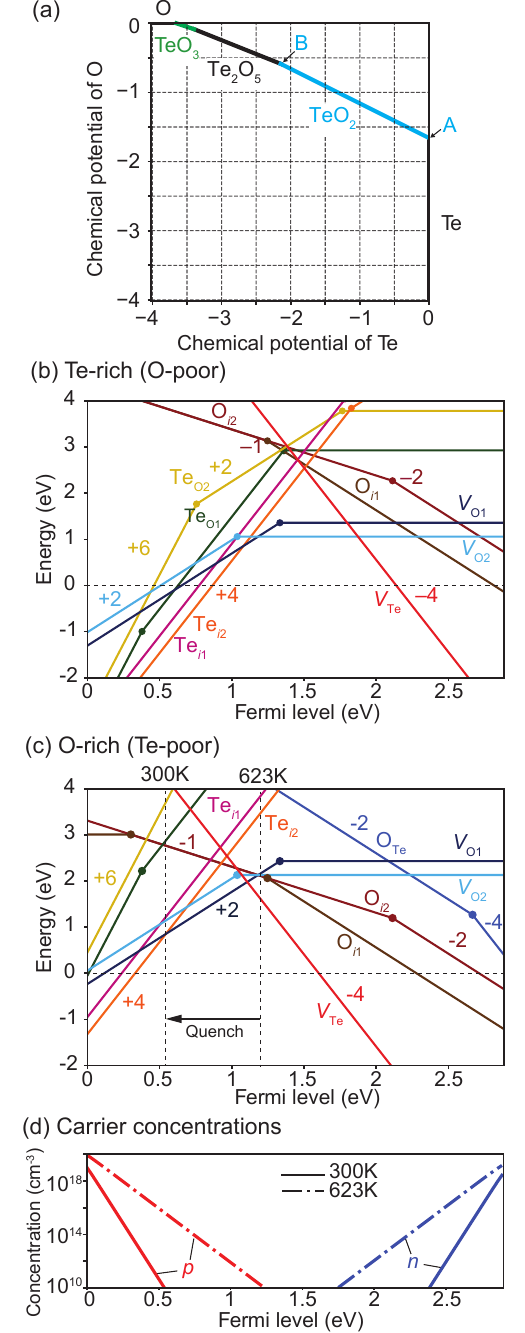}\\
\captionof{figure}{(a) Chemical potential diagram of the Te-O binary system.
    Native defect formation energies
    at the (b) Te-rich and (c) O-rich conditions in $\beta$-TeO$_2$.
    In (c), the equilibrium Fermi levels at 623K and those quenched to 300K are shown in vertical dashed lines.
    Plots over a wider formation energy range are shown in the \sm.
    (d) Carrier concentrations as a function of the Fermi level (see text for details.)}
\label{fig:native}
\end{figure}
\begin{figure}
\centering
\includegraphics[width=8cm]{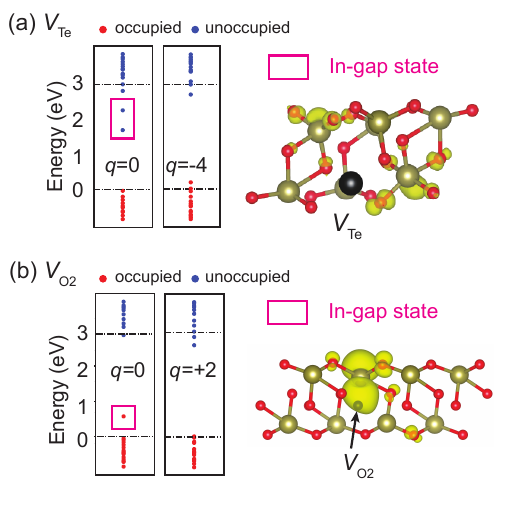}\\
\captionof{figure}{
    Single-particle levels at $k=(1/4, 1/4, 0)$ in the supercell models
    with (a) $V_\text{Te}$ at $q=0$ and $q=-4$, and (b) $V_\text{O2}$ at $q=0$ and $q=+2$.
    The values are alighed to the valence band maximum in the perfect supercell.
    For reference, the horizontal dash-dotted lines denote
    the band edge positions in the perfect supercell.
    The isosurfaces of the squared wavefunctions for the in-gap states,
    enclosed within rectangles
    are also visualized using VESTA~\cite{Momma:db5098}.
    The isosurfaces correspond to 10\% of the maxima in each case.}
\label{fig:native_atomic}
\end{figure}
\begin{table*}
\captionof{table}{
    The calculated Fermi level, defect, and carrier concentrations under
    the O-rich conditions at 623~K and 300~K for $\beta$-TeO$_2$ are
    provided in units of $\text{cm}^{-3}$.
    Details are discussed in the main text.
    When the calculated concentration is lower than 1 $\text{cm}^{-3}$,
    they are indicated as ``--".
}
\label{tab:con}
\begin{tabular}{P{1.0cm}P{1.5cm}P{1.5cm}P{1.5cm}P{1.5cm}P{1.5cm}P{1.5cm}P{1.5cm}P{1.5cm}}
\hline
\hline
$T$ (K)& $E_f$ (eV) & $p$ & $n$ & $\text{V}_{\text{O1}}^{+2}$ & $\text{V}_{\text{O2}}^{+2}$ & $\text{Te}_{\text{i1}}^{+4}$  & $\text{Te}_{\text{i2}}^{+4}$ & $V_{\text{Te}}^{-4}$\\
\hline
623 & 1.21 & 2 $\times$ $10^{10}$ & 5 $\times$ $10^5$ & 5 $\times$ $10^4$ & 2 $\times$ $10^2$ & --  & -- & 4 $\times$ $10^{9}$\\
300 & 0.53 & 2 $\times$ $10^{10}$ & -- & 5 $\times$ $10^4$ & 1 $\times$ $10^5$ & --  & -- & 3 $\times$ $10^{9}$\\
\hline
\hline
\end{tabular}
\end{table*}

%
Since the $\beta$ phase has a lower fundamental gap while maintaining transparency
and a lower hole effective mass, especially in the $a$--$b$ plane,
it should be the most suitable phase as a $p$-type TCO\@. 
Hereinafter, we thus focus on investigating native defects and hole dopability in the $\beta$ phase.
The unit cell of $\beta$-\ce{TeO2} consists of two TeO$_2$ layers, and holds two distinct oxygen sites:
one outside the layer (O1) and the other within the layer (O2) [see Fig.~\ref{fig:3phase}(a)].
In this study, we investigate Te and O vacancies ($V_X$, $X= \text{Te, O1, O2}$), and
Te-on-O and O-on-Te antisites (Te$_\text{O1}$, Te$_\text{O2}$, and O$_\text{Te}$).
We also calculated interstitial sites, the initial positions of which
are at the locations with the lowest ($i1$) and second-lowest ($i2$)
charge density in the unit cell, as shown in Fig.~\ref{fig:3phase}(a).
We calculated Te and O at these two sites ($X_{i1}$ and $X_{i2}$, $X=\text{Te, O}$).

In $\beta$-TeO$_2$, based on the typical oxidation number of O, we would expect
the oxidation states of Te and O to be $+4$ and $-2$, respectively.
We calculated the $q$ from $0$ to $-4$ for $V_\text{Te}$
and from $0$ to $+2$ for $V_\text{O}$
since the closed-shell electronic structures of $V_\text{Te}$ and $V_\text{O}$
are realized at $-4$ and $+2$ charge states, respectively.
In contrast, we calculated the $q$ from $0$ to $+4$ for Te$_i$
and from $0$ to $-2$ for O$_i$ in an opposite manner.
For the substituted defects, we calculated a set of $q$
by regarding them as combinations of vacancies and interstitials.
For example, Te$_\text{O}$ is composed of Te$_i$ and $V_\text{O}$.
Thus, we considered $q$ from $0$ to $+6$ for it.
On the flip side, $q$ from $0$ to $-6$ are considered for O$_\text{Te}$.
Based on their charge states, the candidate donor-type defects are
$V_\text{O}$, $\text{Te}_\text{O}$, and $\text{Te}_i$,
whereas the candidate acceptor-type defects are
$V_{\text{Te}}$, $\text{O}_{\text{Te}}$, and $\text{O}_i$.

The CPD of the Te-O binary system is shown in Fig.~\ref{fig:native}(a).
At the Te-rich (O-rich) condition, \ce{TeO2} is equilibrated with Te (\ce{Te2O5}).
Figures~\ref{fig:native}(b) and (c) show formation energies
of the native defects at the Te-rich and O-rich conditions,
which correspond to vertices A and B in Fig.~\ref{fig:native}(a), respectively.

$V_\text{Te}$ at $q=-4$ makes very deep donor pinning level
even at the Te-rich condition, which may lead to difficulties
for synthesizing the $n$-type $\beta$-\ce{TeO2}.
When we plot the eigenvalues of $V_\text{Te}^{-4}$,
any obvious in-gap states are not found,
whereas there are two unoccupied states in the band gap at $q=0$.
Consequently, the stable charge state of $V_\text{Te}$
transits from $0$ to $-4$, depending on the Fermi level
(see also the Supplementary Material).
The unoccupied in-gap states at $q=0$ are
composed of the neighboring
Te-$5s$ and -$5p$ orbitals hybridized with the O-$2p$ orbitals,
as shown in Fig.~\ref{fig:native_atomic}(a).

In contrast, $V_{\text{O1}}$ creates a relatively
shallower acceptor pinning level at around 0.1 eV
at the O-rich condition.
On the other hand, $V_{\text{O2}}$ has a slightly higher energy,
resulting in absence of the pinning level.
The $U^\text{eff}(0/+1/+2)$ for $V_{\text{O1}}$ and $V_{\text{O2}}$
are $-0.33$ and $-0.46$ eV, respectively.
Such negative $U$ behavior for oxygen vacancies
has been commonly reported in representative oxides
like ZnO~\cite{PhysRevB.77.245202} and MgO~\cite{tanaka2002theoretical}.
However, in contrast to the electride-like F-center states reported in ZnO and MgO,
two (one) electrons at $q=0$ ($q=+1$)
are captured by the neighboring Te atom in $\beta$-\ce{TeO2}
[see Fig.~\ref{fig:native_atomic}(b)].

Among the other native defects,
only $\text{Te}_{i}$ take relatively lower formation energies.
Especially, $\text{Te}_{i2}$ pins the Fermi level at 0.34 eV
even at the O-rich condition, setting a lower limit for the Fermi level,
where $10^{13}$ cm$^{-3}$ holes are introduced at 300~K
as shown in Fig.~\ref{fig:native}(d).
Such hole concentration is insufficient for semiconductor applications.
The chemical potentials at the O-rich condition are
determined from the total energies of \ce{Te2O5} and \ce{TeO2}.
However, the Te chemical potential might be further decreased to suppress Te$_i$
when the sample is synthesized, for instance, at high oxygen partial pressure,
with suppressing the competing \ce{Te2O5} and \ce{TeO3}.

Figure~\ref{fig:native}(c) and (d) and Table~\ref{tab:con}
present the Fermi level, as well as the defect and carrier concentrations,
based on the charge neutrality condition~\cite{PhysRevB.90.125202} at the O-rich condition.
The temperature is then set to 623~K,
following the growth temperature in the experiment~\cite{shi2023electronic}.
The equilibrium Fermi level is primarily determined by $\text{V}_\text{O}^{+2}$ and $V_\text{Te}^{-4}$.
However, their formation energies are so high that their concentrations are extremely low.
Consequently, the Fermi level is located in the middle of the band gap,
resulting in an intrinsically insulating behavior.

We also calculated the Fermi level at 300~K assuming
that the defect concentrations are quenched from 623~K,
with taking different charge states
in the ratio determined by the Boltzmann distribution~\cite{PhysRevB.90.125202}.
After the quench, the Fermi level is largely shifted from 1.21 eV to 0.53 eV.
Since the Fermi level does not cross the transition levels of the dominant defects, \ie, $V_\text{O1}$ and
$V_\text{Te}$, the hole carrier and dominant defect concentrations
remain unchanged during the quench.
\subsection{$p$-type dopability of $\beta$-TeO$_2$}\label{sec:dopants}
%
\begin{figure*}
\centering
\includegraphics[width=16cm]{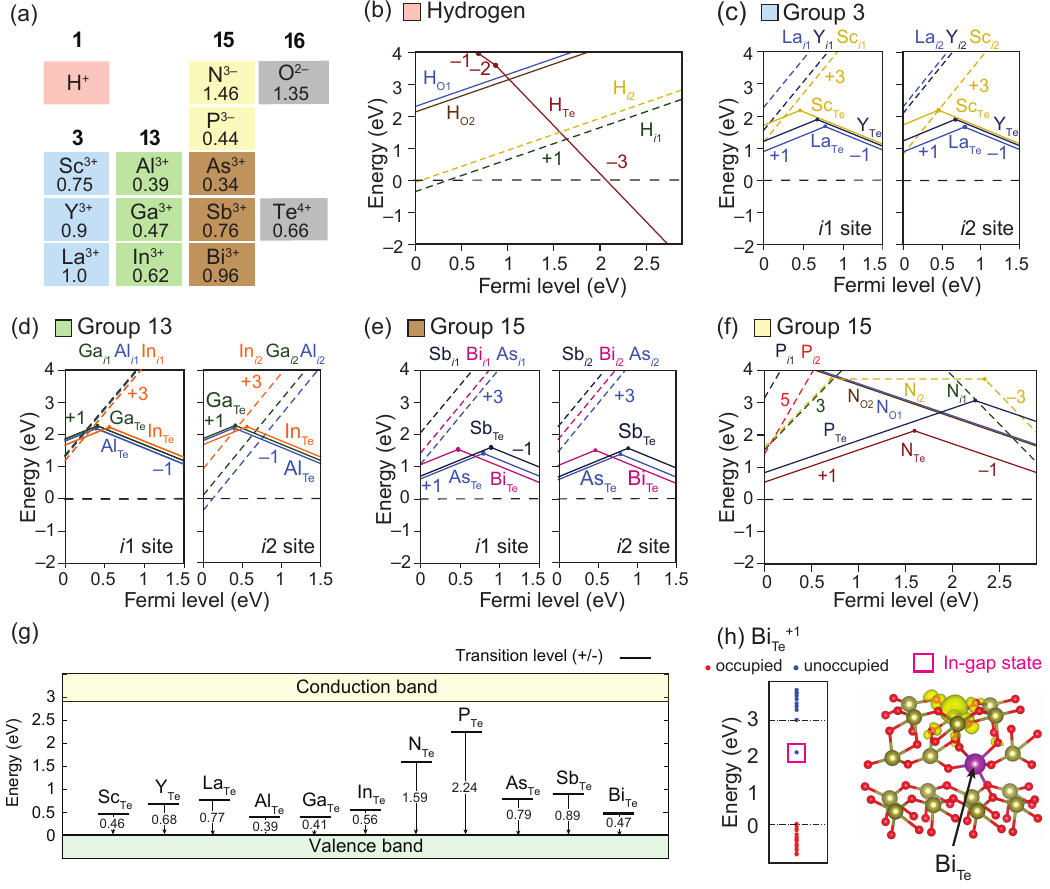}\\
\captionof{figure}{
    (a) Considered elements as dopants in this work.
    Their charge states that are expected in $\beta$-\ce{TeO2} as well as the ionic radii
    in units of $\mathrm{\AA}$~\cite{shannon1976revised} are also shown.
    Information on Te and O is also shown for reference.
    (b)--(f) Formation energies of the impurities at the O-rich conditions.
    Solid lines indicate the formation energies of the substitution-type defects,
    whereas dashed lines represent those of the interstitials.
    Plots over a wider energy range are shown in the \sm.
    (g) Transition levels of the 11 dopants at the Te substituted site.
    (h) Same as Fig.~\ref{fig:native_atomic} but for Bi$_\text{Te}^{+1}$.
    The isosurfaces correspond to 5\% of the maximum.}
\label{fig:defect}
\end{figure*}
%

Since $\beta$-\ce{TeO2} is intrinsically insulating even at the O-rich condition,
we consider 11 elements as acceptor dopants
whose oxidation states and ionic radii are shown in Fig.~\ref{fig:defect}(a).
In addition, we considered the hydrogen impurity, as it is the most ubiquitous impurity in semiconductors and insulators~\cite{VandeWalle2003}.
Group 3 and 13 elements (Sc, Y, La, Al, Ga, and In) generally take $+3$ oxidation states
and are expected to introduce holes while substituting Te.
The N and P in group 15 are considered to act as acceptors by placing at the O site with their $-3$ oxidation states.
As, Sb, and Bi are also expected to behave as acceptors
at both Te and O sites due to their amphoteric behaviors by taking $+3$ or $-3$ oxidation states.
The calculated formation energies of the dopants at the O-rich conditions
are shown in Figs.~\ref{fig:defect}(b)--(f).
The calculated competing phases composed of the impurity elements at the O-rich condition are tabulated in the \sm.

As shown in Fig.~\ref{fig:defect}(b), the hydrogen interstitial
shows a Fermi level pinning at around 0.4 eV similar with Te$_{i2}$ [see Fig.~\ref{fig:native}(c)],
which is not beneficial to the $p$-type doping.
Thus, the hydrogen partial pressure should be minimized through, \textit{e.g.},
post-process annealing.
In turn, H$_\text{O}$ defects act as donors at any Fermi level within the band gap,
but their energies are considerably high at the O-rich condition.

Calculation results of group 3 (Sc, Y, La), 13 (Al, Ga, In), and 15 (N, P, As, Sb, Bi) elements
are shown in Figs.~\ref{fig:defect}(c)--(f).
When they are substituted to Te,
they behave as acceptors, as expected from their $+3$ oxidation states.
Among As, Sb, and Bi, Bi$_\text{Te}$ shows a shallower acceptor level.
N$_\text{Te}$ and P$_\text{Te}$ exhibit the lowest formation energies with amphoteric behaviors,
similar to the other elements considered, but their transition levels are much higher (1.59 and 2.24 eV, respectively).
It should be noted that they exhibit a \textit{negative} $U$ behavior in common,
where $q=-1$ and $+1$ charge states become more stable over the $q=0$ states for the whole range of the Fermi level.
The $U^\text{eff}(-1/0/+1)$ widely range between $-0.07$ and $-1.88$ eV
(see the \sm\ for their values).

The $(+/-)$ transitions are ranged in 0.39 -- 0.77 eV above the VBM [see Fig.~\ref{fig:defect}(g)],
meaning the introduced holes are tightly bound compared to the shallow hydrogentic states,
where the simple model estimates the acceptors level around 40 meV from the spherically averaged dielectric constant and effective mass shown in Table.~\ref{tab:lat}~\cite{peter2010fundamentals}.
The electronic structure of Bi$_\text{Te}^{+1}$ is shown in Fig.~\ref{fig:defect}(h).
When introducing a Bi at $q=+1$, two holes are captured by a lone pair state at the neighboring Te.
We also calculated the two localized holes without Bi, namely
we replace Bi with Te and change the number of electrons to retain the two holes.
Consequently, we have found such localized holes are not stabilized energetically
without help of attractive electrostatic interaction from a negatively charged Bi$_\text{Te}^{-1}$.
In other words, the self-trapped holes are unstable in $\beta$-\ce{TeO2}.

The formation energies of the interstitials are also shown in Figs.~\ref{fig:defect}(b)--(f).
Since interstitials exhibit positive charge states, they compensate hole carriers.
The formation energy of the interstitial is generally lower as the ionic radius decreases.
Thus, dopants with relatively large ionic radii could be more preferential for the acceptor doping.
An exception is Bi, which shows a lower formation energy than Sb, though it has a larger ionic radius.
Although the transition level of Bi$_\text{Te}$ is slightly higher than that of Al$_\text{Te}$,
the interstitial formation energies are high enough not to self-compensate.
Therefore, Bi could be the best acceptor dopant.

The transition level of Bi$_{\text{Te}}$ is still located at 0.47 eV, close to the equilibrium Fermi level at the O-rich condition at 300K [see Fig.~\ref{fig:native}(c)],
at which the hole concentration is estimated to be around $10^{11} \text{cm}^{-3}$ [Fig.~\ref{fig:native}(d)].
Since successful fabrication of $p$-type FET with $\alpha$-TeO$_2$ has been reported \cite{Devabharathi2024}, 
we also calculated Bi$_{\text{Te}}$ in $\alpha$-TeO$_2$ with the same $\alpha^\text{HSE}$.
The acceptor level of Bi$_{\text{Te}}$ is found at around 1.0 eV,
which is about 0.5 eV deeper than the acceptor level in $\beta$-\ce{TeO2} (see the \sm).
This implies that FET devices should be realized not only with
the 2D $\beta$ phase \cite{zavabeti2021high} but also with the 3D $\beta$ phase in experiments.

In addition, $\alpha^\text{HSE}$ is adjusted to match the optical band gap rather than the band edges in this study.
This implies that there is a possibility that  the actual VBM is higher in energy, leading to shallower acceptor levels than our calculations.
Furthermore, when forming acceptor-type impurity levels near the VBM with considerable dopant concentration,
it is possible to form an impurity band that can merge
with the valence band and provide mobile hole carriers~\cite{samarth2012battle,wang2019excitation}.
Therefore, we suggest introducing Bi dopant at high concentrations
under O-rich and H-poor conditions in experiments.

\section{Conclusions}\label{sec:conclu}
In this study, we have investigated the potential of $\beta$-\ce{TeO2} as a new transparent conducting oxide.
Based on our calculations using a band-gap-tuned hybrid functional,
it exhibits transparency but a smaller fundamental gap than the optical gap,
which is advantageous for carrier doping.
We have also found the in-plane hole effective mass is only 0.74--0.75 $m_0$, indicating its good hole conductivity.

Our investigation into the native defects of $\beta$-\ce{TeO2} shows that
it is intrinsically insulating regardless of the growth condition.
Furthermore, we evaluated the formation energies and transition levels of 11 acceptor dopants and a hydrogen impurity.
We found that although none of the acceptor dopants exhibit hydrogenic states,
Sc, Al, Ga, and Bi show relatively shallower acceptor levels when substituted for the Te site.
The absence of shallow hydrogenic states is attributed to the
stabilization of the localized holes by the attractive electrostatic interaction from the acceptor dopants.
Considering the formation energies of various types of defects,
we suggest that $p$-type doping may be achieved
by doping Bi at a high concentration under O-rich and H-poor conditions.

\section*{Acknowledgement}
This research has been financially supported by JSPS KAKENHI Grant Number 22H01755 and 23KF0030
and the E-IMR project at IMR, Tohoku University.
Part of calculations were conducted using MASAMUNE-IMR (Project No. 202312-SCKXX-0408) and ISSP supercomputers.

%

\end{document}


\title{Supplemental material: Native defects and $p$-type dopability in transparent $\beta$-TeO$_2$: A first-principles study}

\author{Vu Thi Ngoc Huyen}
\author{Soungmin Bae}
\author{Rafael Costa-Amaral}
\author{Yu Kumagai}
\thanks{yukumagai@tohoku.ac.jp}
\affiliation{Institute for Materials Research, Tohoku University, Sendai, Miyagi 980--8577, Japan}

\maketitle
\clearpage

\begin{table}[h!]
\caption{Information on the PAW data sets adopted in this study.}
\label{tab:paw_data_sets}
\centering
\begin{tabular}{cccc}
\hline
\hline
Element & VASP symbol & Valence orbitals & PAW core radii (Å) \\
\hline
O & O & (2p)$^4$ (2s)$^2$ & 0.80 \\
Te & Te & (5p)$^4$ (5s)$^2$ & 1.22 \\
H   & H &  (1s)$^1$ & 0.58 \\
Sc & Sc & (3p)$^6$ (4s)$^1$ (3d)$^2$ & 1.59 \\
Y  & Y\_sv & (4p)$^6$ (5s)$^2$ (4d)$^1$ & 1.48 \\
La   & La & (5p)$^6$ (6s)$^2$ (5d)$^1$& 1.48 \\
Al    & Al & (3p)$^1$ (3s)$^2$ & 1.00 \\
Ga     & Ga\_d & (4p)$^1$ (4s)$^2$ & 1.22 \\
In      & In & (5p)$^1$ (5s)$^2$ & 1.64 \\
N       & N & (2p)$^3$ (2s)$^2$ & 0.79 \\
P & P & (3p)$^3$ (3s)$^2$ & 1.00\\
As & As & (4p)$^3$ (4s)$^2$ & 1.11\\
Sb & Sb & (5p)$^3$ (5s)$^2$ & 1.22\\
Bi      & Bi & (6p)$^3$ (6s)$^2$ & 1.59 \\
Sn      & Sn & (5p)$^2$ (5s)$^2$ & 1.59 \\
\hline
\hline
\end{tabular}
\end{table}


\begin{table}
\caption{Chemical potentials ($\Delta \mu_{i}$, where $i$ means the impurity elements) of the ternary systems that including each impurity at point A in Fig.~3(a).
Tha values are relative to those of the standard states, namely simple substances or molecules.
The competing phases that include impurity elements are described in the second row.
}
\label{tab:chem_pot_A}
\centering
\begin{tabular}{ccccccccccccc}
\hline
\hline
 & Al & Bi & Ga & H & In & La & Sc & Y & N & P & As & Sb \\
\hline
$\Delta \mu_{i}$ (eV) & -5.68 & -0.97 & -2.65 & -0.76 & -2.11 & -7.60 & -7.24 & -7.87 & 0 & -4.31 & -0.69 & -1.17 \\
Competing phases  & Al$_2$O$_3$ & Bi$_2$Te$_4$O$_{11}$& Ga$_2$O$_3$ & Te(HO)$_6$ & In$_2$(TeO$_3$)$_3$ & La$_2$Te$_4$O$_{11}$ &Sc$_2$Te$_5$O$_{13}$ & Y$_2$Te$_5$O$_{13}$ & N$_2$ & Te$_4$P$_2$O$_{13}$ & As$_2$O$_3$ & SbO$_2$ \\
\hline
\hline
\end{tabular}
\end{table}

\begin{table}
    \caption{Same as Table.~\ref{tab:chem_pot_A} but at point B in Fig.~3(a).}
    \label{tab:chem_pot_B}
    \centering
    \begin{tabular}{ccccccccccccc}
        \hline
        \hline
 & Al & Bi & Ga & H & In & La & Sc & Y & N & P & As & Sb \\
        \hline
     $\Delta \mu_{i}$ (eV) & -7.29 & -2.94 & -4.67 & -1.51 & -3.98 & -9.38 & -9.20 & -9.66 & 0 & -6.99 & -3.03 & -3.68 \\
        Competing phases  & Al$_2$O$_3$& BiTeO$_4$ & Ga$_2$TeO$_6$ & TeHO$_3$ & In$_2$TeO$_6$ & La$_2$TeO$_6$ & Sc$_2$TeO$_6$ & Y$_2$TeO$_6$ & N$_2$ & Te$_4$P$_2$O$_{13}$ & Te$_3$As$_2$O$_{11}$ & Sb$_2$Te$_2$O$_9$ \\
        \hline
        \hline
    \end{tabular}
\end{table}
%
%

\begin{table}
\caption{Considered competing phases}
\label{tab:competing}
\begin{tabular}{ c c }
\hline
\hline
   & Competing phases   \\
\hline
$\beta$-TeO$_2$ & O$_{2}$(g), Te, Te$_2$O$_5$, Te$_2$O$_3$, $\beta$-TeO$_2$ \\
H & H$_{2}$(g), H$_2$O(g), Te(HO)$_6$, TeHO$_3$\\
Sc & Sc, Sc$_2$O$_3$, Sc$_2$Te, Sc$_2$Te$_3$,  Sc$_2$Te$_5$O$_{13}$, Sc$_2$TeO$_6$, Sc$_9$Te$_2$, ScTe \\
Y & Y, Y$_2$O$_3$, Y$_2$Te$_5$O$_{13}$, Y$_2$TeO$_2$, Y$_2$TeO$_6$, Y$_6$TeO$_{12}$,YTe, YTe$_3$,Y$_2$Te$_3$ \\
La & La, La$_2$O$_3$, La$_2$Te$_3$, La$_2$Te$_4$O$_{11}$,  La$_2$TeO$_2$, La$_2$TeO$_6$, La$_3$Te$_4$, LaTe, LaTe$_2$, LaTe$_3$\\
Al & Al, Al$_2$O$_3$, Al$_2$Te$_3$, Al$_2$Te$_5$ \\
Ga & Ga, Ga$_2$O$_3$, Ga$_2$Te$_3$, Ga$_2$Te$_5$, Ga$_2$TeO$_6$, Ga$_7$Te$_{10}$, GaTe \\
In & In, In$_2$(TeO$_3$)$_3$, In$_2$O$_3$, In$_2$TeO$_6$, In$_4$Te$_3$, In$_7$Te$_{10}$ \\
N &  N$_{2}$(g), NO$_2$(g)\\
P & P, P$_2$O$_5$, Te$_2$P$_2$O$_9$, Te$_3$P$_2$O$_{11}$, Te$_4$P$_2$O$_{13}$ \\
As & As, Te$_3$As$_2$, As$_2$O$_3$, As$_2$O$_5$, Te$_3$As$_2$O$_{11}$ \\
Sb & Sb, Sb$_2$O$_3$, Sb$_2$O$_5$, Sb$_2$Te$_3$, SbO$_2$, SbTe$_2$O$_9$ \\
Bi & Bi, Bi$_2$O$_3$, Bi$_2$Te$_3$, Bi$_2$Te$_4$O$_{11}$, Bi$_2$TeO$_{2}$, Bi$_2$TeO$_{5}$, Bi$_2$TeO$_{6}$, Bi$_4$O$_7$, BiO$_2$, BiTeO$_4$ \\
\hline
\hline
\end{tabular}
\end{table}
\begin{table}
\caption{
    The calculated $U^\text{eff}(-1/0/+1)$ for the substituted impurities in $\beta$-TeO$_2$.}
\label{tab:uvalue}
\begin{tabular}{cc}
\hline
\hline
$\text{Sc}_{\text{Te}}$ & $-0.56$ \\
$\text{Y}_{\text{Te}}$ & $-0.60$ \\
$\text{La}_{\text{Te}}$ & $-0.89$ \\
$\text{Al}_{\text{Te}}$ & $-0.52$ \\
$\text{Ga}_{\text{Te}}$ & $-0.07$ \\
$\text{In}_{\text{Te}}$ & $-0.73$ \\
$\text{N}_{\text{Te}}$ & $-1.88$ \\
$\text{P}_{\text{Te}}$ & $-1.21$ \\
$\text{As}_{\text{Te}}$ & $-0.58$ \\
$\text{Sb}_{\text{Te}}$ & $-0.37$ \\
$\text{Bi}_{\text{Te}}$ & $-0.36$ \\
\hline
\hline
\end{tabular}
\end{table}
\begin{figure*} 
\centering 
\includegraphics[width=18cm]{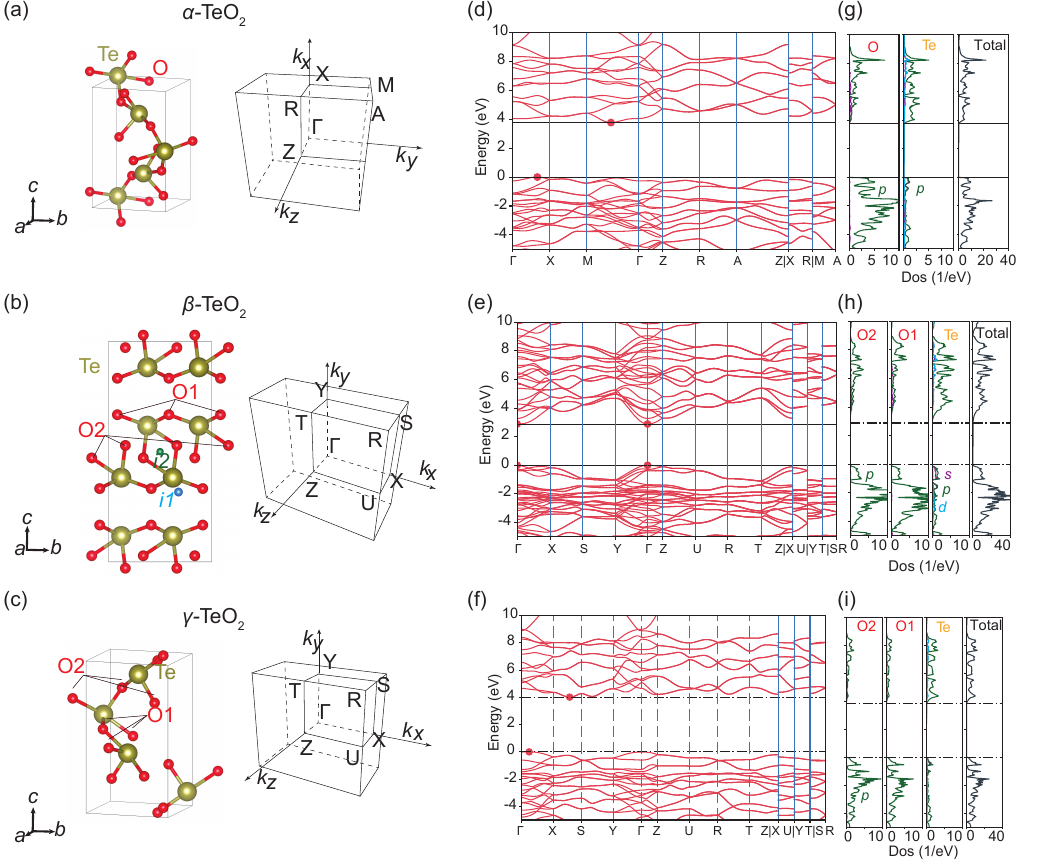}\\
\caption{Crystal structures, first Brillouin zones, electronic band structures, and densities of states
for (a) $\alpha$-, (b) $\beta$-, and (c) $\gamma$-TeO$_2$.}
\label{fig:spolymorphous}
\end{figure*}
\begin{figure*} 
\centering 
\includegraphics[width=18cm]{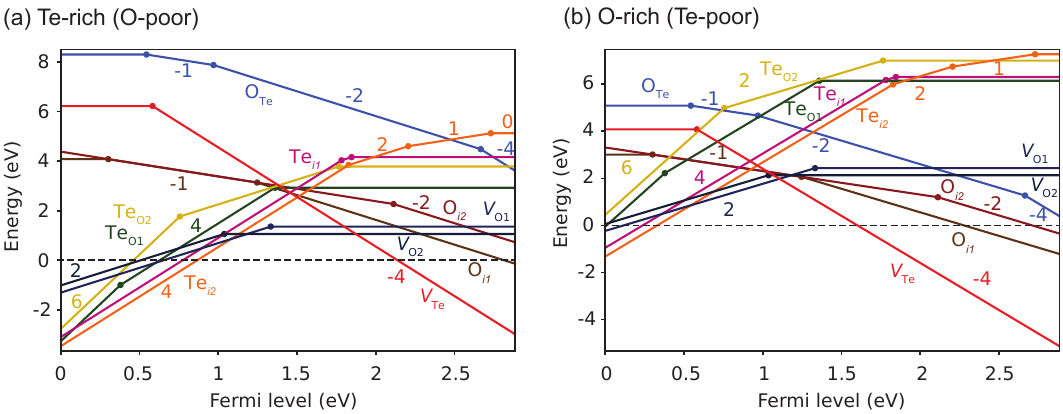}\\
\caption{Same as Figs.~3(b) and (c) in the main text but with expanded $y$-axis ranges.}
\label{fig:snative}
\end{figure*}
\begin{figure*} 
\centering 
\includegraphics[width=18cm]{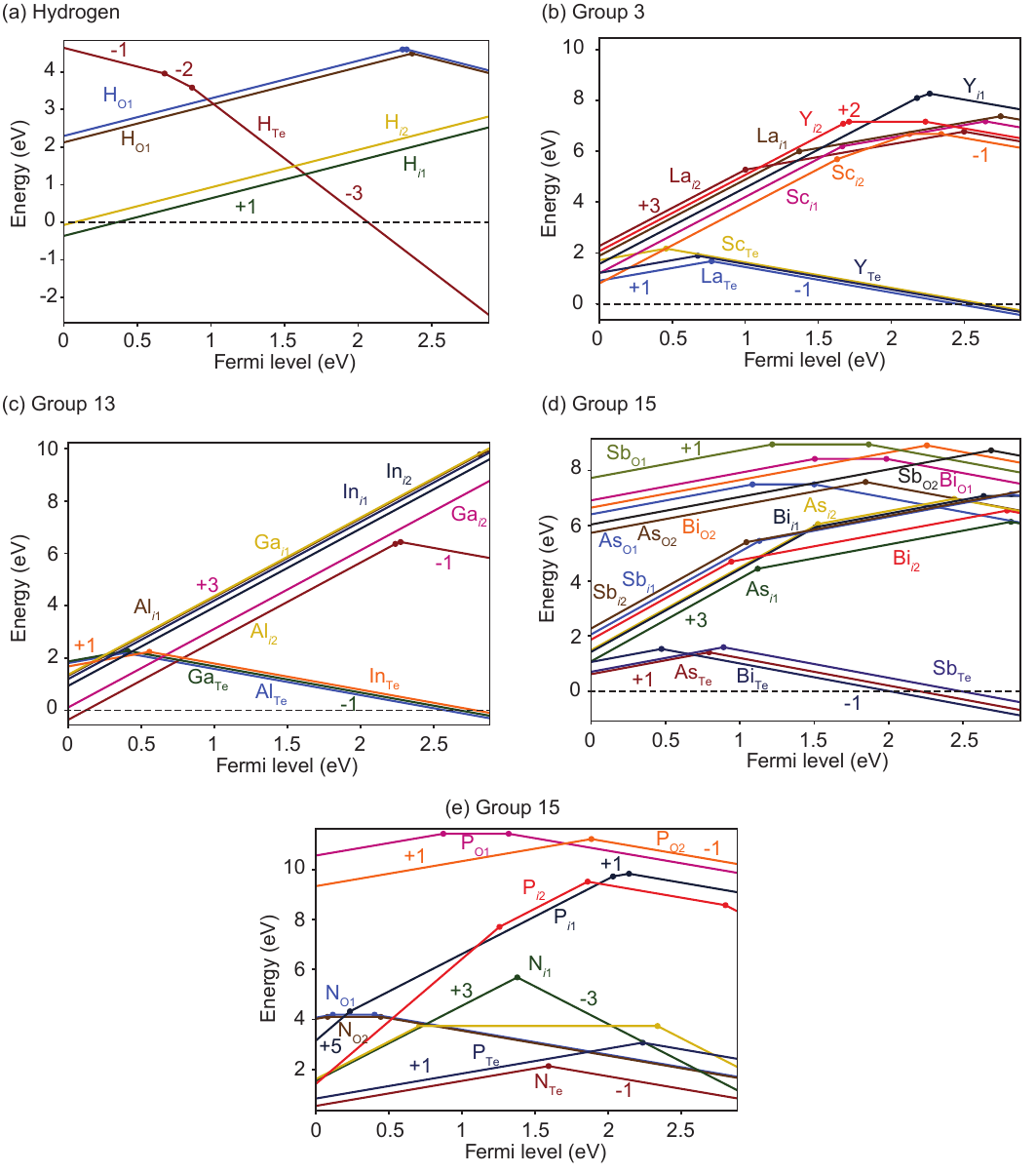}\\
\caption{Same as Figs.~5(b)--(f) in the main text but with expanded $y$-axis ranges.}
\label{fig:sdefect}
\end{figure*}

\begin{figure} 
\centering 
\includegraphics[width=8cm]{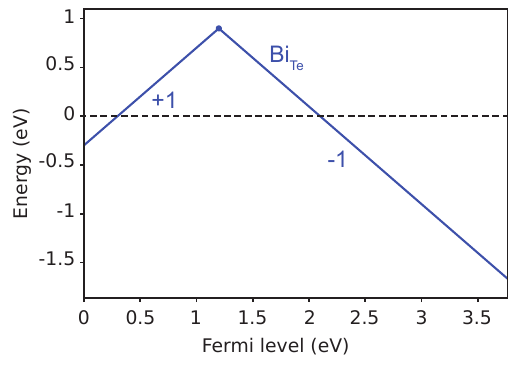}\\
    \caption{Defect formation energies of the Bi-on-Te substitution-type defect in $\alpha$-TeO$_2$.}
    \label{fig:Bi_Te_in_alpha}
\end{figure}
%
\bibliography{teo2.bib}